\documentclass[10pt,superscriptaddress,tightenlines,twocolumn,amsmath,amssymb,aps, pra]{revtex4-2}
\usepackage[
a4paper,
left=13.5mm,     
right=13.5mm,
top=25mm,
bottom=25mm,
columnsep=5mm,    
headheight=15pt
]{geometry}

\usepackage[T1]{fontenc}
\usepackage[utf8]{inputenc}
\usepackage{lmodern}
\usepackage{CJKutf8}
\usepackage{mathrsfs}
\UseRawInputEncoding
\usepackage{graphicx}
\usepackage{dcolumn}
\usepackage{bbm}
\usepackage{color}
\usepackage{amsmath}
\usepackage{amssymb}
\usepackage{amsthm}
\usepackage{subfigure}
\usepackage{braket}
\usepackage{indentfirst}
\usepackage{mathrsfs}
\usepackage[linesnumbered,ruled,vlined]{algorithm2e}
\usepackage{enumerate}
\usepackage{enumitem}
\usepackage{algorithmic}
\usepackage{amsfonts}
\usepackage[normalem]{ulem} 

\usepackage{makecell}
\usepackage{lmodern}
\usepackage{lineno}
\usepackage{multirow}

\usepackage{newtxtext,newtxmath}

\usepackage{graphicx}
\usepackage{tabularx}
\usepackage{ulem}
\usepackage[backref=false,
           colorlinks=true,
           linkcolor=blue,
           citecolor=blue]{hyperref}
\usepackage{physics}
\usepackage{braket}
\usepackage{hyperref}
\usepackage{url}
\usepackage{booktabs}
\usepackage{multirow}
\usepackage{makecell} 
\usepackage{setspace}

\newcommand{\be}{\begin{equation}}
\newcommand{\ee}{\end{equation}}

\newcommand{\RNum}[1]{\uppercase\expandafter{\romannumeral #1\relax}}

\makeatletter

\newcommand{\Rmnum}[1]{\expandafter\@slowromancap\romannumeral #1@}

\makeatother

\usepackage{xcolor}
\colorlet{RED}{red}
\colorlet{BLACK}{black}

\providecommand{\U}[1]{\protect\rule{.1in}{.1in}}

\hypersetup{colorlinks=true, linkcolor=red, citecolor=blue, urlcolor=black}

\makeatletter
\newcommand{\newreptheorem}[2]{\newtheorem*{rep@#1}{\rep@title}\newenvironment{rep#1}[1]{\def\rep@title{#2 \ref*{##1}}\begin{rep@#1}}{\end{rep@#1}}}
\makeatother

\newreptheorem{theorem}{Theorem}

\newreptheorem{lemma}{Lemma}

\newreptheorem{proposition}{Proposition}

\begin{document}
\title{A fault-tolerant quantum blockchain deployed on commercial telecommunications network}

\author{Yongqiang Du}
\thanks{These authors contributed equally to this work.}
\affiliation{Guangxi Key Laboratory for Relativistic Astrophysics, School of Physical Science and Technology, Guangxi University, Nanning 530004, China}

\author{Chen-Xun Weng}
\thanks{These authors contributed equally to this work.}
\affiliation{National Laboratory of Solid State Microstructures and School of Physics, Collaborative Innovation Center of Advanced Microstructures, Nanjing University, Nanjing 210093, China}
\affiliation{School of Physics and Key Laboratory of Quantum State Construction and Manipulation (Ministry of Education), Renmin University of China, Beijing 100872, China}

\author{Feng Xie}
\thanks{These authors contributed equally to this work.}
\affiliation{Guangxi Key Laboratory for Relativistic Astrophysics, School of Physical Science and Technology, Guangxi University, Nanning 530004, China}

\author{Ming-Yang Li}
\thanks{These authors contributed equally to this work.}
\affiliation{National Laboratory of Solid State Microstructures and School of Physics, Collaborative Innovation Center of Advanced Microstructures, Nanjing University, Nanjing 210093, China}
\affiliation{School of Physics and Key Laboratory of Quantum State Construction and Manipulation (Ministry of Education), Renmin University of China, Beijing 100872, China}

\author{Mingxuan Zhang}
\affiliation{Guangxi Key Laboratory for Relativistic Astrophysics, School of Physical Science and Technology, Guangxi University, Nanning 530004, China}

\author{Xin Hua}
\affiliation{National Information Optoelectronics Innovation Center (NOEIC), Wuhan 430074, China}

\author{Xin An}
\affiliation{Guangxi Key Laboratory for Relativistic Astrophysics, School of Physical Science and Technology, Guangxi University, Nanning 530004, China}

\author{Xiang Guan}
\affiliation{Guangxi Key Laboratory for Relativistic Astrophysics, School of Physical Science and Technology, Guangxi University, Nanning 530004, China}

\author{Xin Liu}
\affiliation{Guangxi Key Laboratory for Relativistic Astrophysics, School of Physical Science and Technology, Guangxi University, Nanning 530004, China}

\author{Zhenrong Zhang}
\affiliation{Guangxi Key Laboratory of Multimedia Communications and Network Technology, School of Computer, Electronics, and Information, Guangxi University, Nanning 530004, China}

\author{Xi Xiao}\email{xiaoxi@noeic.com}
\affiliation{National Information Optoelectronics Innovation Center (NOEIC), Wuhan 430074, China}

\author{Hua-Lei Yin}\email{hlyin@ruc.edu.cn}
\affiliation{School of Physics and Key Laboratory of Quantum State Construction and Manipulation (Ministry of Education), Renmin University of China, Beijing 100872, China}
\affiliation{National Laboratory of Solid State Microstructures and School of Physics, Collaborative Innovation Center of Advanced Microstructures, Nanjing University, Nanjing 210093, China}

\author{Kejin Wei}\email{kjwei@gxu.edu.cn}
\affiliation{Guangxi Key Laboratory for Relativistic Astrophysics, School of Physical Science and Technology, Guangxi University, Nanning 530004, China}

\date{\today}

\begin{abstract}
\noindent
Popularized by the Bitcoin cryptocurrency, blockchain technology establishes a decentralized digital framework that utilizes cryptographic and consensus protocols to secure data against unauthorized modification. Consequently, blockchain has found broad adoption across diverse fields, including finance, data management, healthcare, and digital asset governance. In the quantum computing era, a paramount objective for blockchain is to preserve its foundational advantages of cryptographic integrity and decentralized fault-tolerant resilience. In principle, quantum digital signatures and quantum Byzantine agreement protocols offer foundational security guarantees and tolerate up to one-half of malicious nodes for blockchain. However, the practical realization of such a quantum-enhanced blockchain remains a significant and multifaceted challenge.
Here, we propose and experimentally demonstrate a fully operational hybrid quantum blockchain architecture built on photonic integrated circuits and deployed over commercially available classical telecommunications infrastructure. 
The system achieves a fault tolerance of nearly one-half, surpassing the classical limit, while reaching consensus on a timescale of seconds.
A deployed food traceability application validates the practicality of the proposed architecture, achieving a throughput of approximately 500 transactions per second. This work establishes a foundation for practical quantum blockchains, enabling secure, scalable, and decentralized information processing in the emerging quantum era.

\end{abstract}

\maketitle


\section{Introduction}
The global market for commercial blockchain technology currently exceeds 10 billion and is projected to expand significantly to over 150 billion in the coming ten years. The cryptocurrency market, a primary application of blockchain technology, has reached a total valuation of approximately 3 trillion, with Bitcoin alone accounting for over half of this total market capitalization~\cite{nakamoto2008peer}. If the foundational security promised by cryptographic and consensus protocols of Bitcoin transactions were to be fundamentally compromised, the repercussions would be severe, eroding trust in the entire cryptocurrency ecosystem and potentially triggering widespread financial instability and economic disruption~\cite{swan2015blockchain,2019-Belotti,2025-Najmus}. 

On one hand, the security of cryptographic systems relies on computational hardness assumptions, such as the difficulty of factoring large integers or solving discrete logarithms~\cite{rivest1978method,miller1985use}. Indeed, studies have indicated that the classical cryptography currently securing Bitcoin is vulnerable to being broken by sufficiently advanced quantum computers~\cite{aggarwal2018quantum,2018-fedorov}. Advances in quantum computing, moving beyond supremacy experiments~\cite{2019-arute,2020-Zhong,madsen2022quantum,gao2025establishing,google2025observation} to milestones in error correction~\cite{google2025quantum,He2025Experimental}, have significantly intensified and brought closer the long-envisioned threat to existing cryptographic systems, rendering it an imminent and critical security challenge. On the other hand, classical distributed networks rely on pairwise communication channels, a structure that inherently bounds consensus protocols like Byzantine agreement~\cite{1982-Lamport,1999-Miguel} to a one-third fault tolerance threshold~\cite{1986-dolev,1986-Fischer}. Besides, even by using pairwise quantum channels, this fundamental limitation cannot be broken~\cite{fitzi2001advances,kiktenko2018quantum}.

By leveraging quantum correlations, fully quantum blockchains are expected to achieve not only information-theoretic security but also overcome the fundamental one-third fault tolerance limitation in quantum internet~\cite{wehner2018quantum}. Significant hurdles persist in both the protocol architecture~\cite{fitzi2001quantum,gaertner2008experimental} and the physical hardware~\cite{2024-Yang-BC,azuma2023quantum,li2024asynchronous} required for a viable, fully quantum blockchains, leaving the technology in an immature state. Fortunately, hybrid quantum blockchains combine classical and quantum components, offering a near-term route to practical implementation~\cite{2024-Yang-BC}.  These architectures allow practical quantum advantages to be realized in blockchains using current state-of-the-art technology, delivering improved security~\cite{2023-Yin,2025-Du} and higher fault tolerance~\cite{2023-Weng,2024-Jing}. Despite these advances, a fully operational hybrid quantum blockchain architecture, which preserves quantum advantages, integrates seamlessly with existing and emerging classical telecommunications infrastructure, and achieves meaningful application throughput for real-world tasks, has not yet been realized.

In this article, we address this critical gap by proposing and experimentally demonstrating a four-layer hybrid quantum blockchain architecture (see Fig.~\ref{fig:architecture}). The quantum layer, comprising a quantum physical layer and a quantum consensus layer, employs a quantum cryptographic protocol~\cite{2023-Yin}  and a quantum Byzantine agreement (QBA) protocol~\cite{2023-Weng}, respectively, preserving quantum advantages of higher fault tolerance and enhanced security. In contrast, the classical layer, consisting of an application layer and a classical logical layer, enables seamless integration of quantum advantages with existing classical infrastructure for data transfer and synchronization.

In the demonstration, we deploy a five-user network built on photonic integrated circuits and operated over commercially available classical telecommunications infrastructure. By employing a prepare-and-measure quantum key generation protocol (KGP) and a recursive QBA, our system achieves fault tolerance approaching one-half, surpassing the classical one-third bound, while maintaining unconditional security against quantum adversaries. We achieve secure key rates of up to 273.49 kbps and consensus rates of 0.43 transactions per second for 1-Mbit messages under realistic conditions. To further illustrate the practicality of the platform, we implement a food traceability application, realizing end-to-end data acquisition, block recording, and querying with a throughput of 500 transactions per second and zero failed transactions.

\section{Architecture}

\begin{figure}[t!]
	\centering
    
		\includegraphics[width=0.5\textwidth]{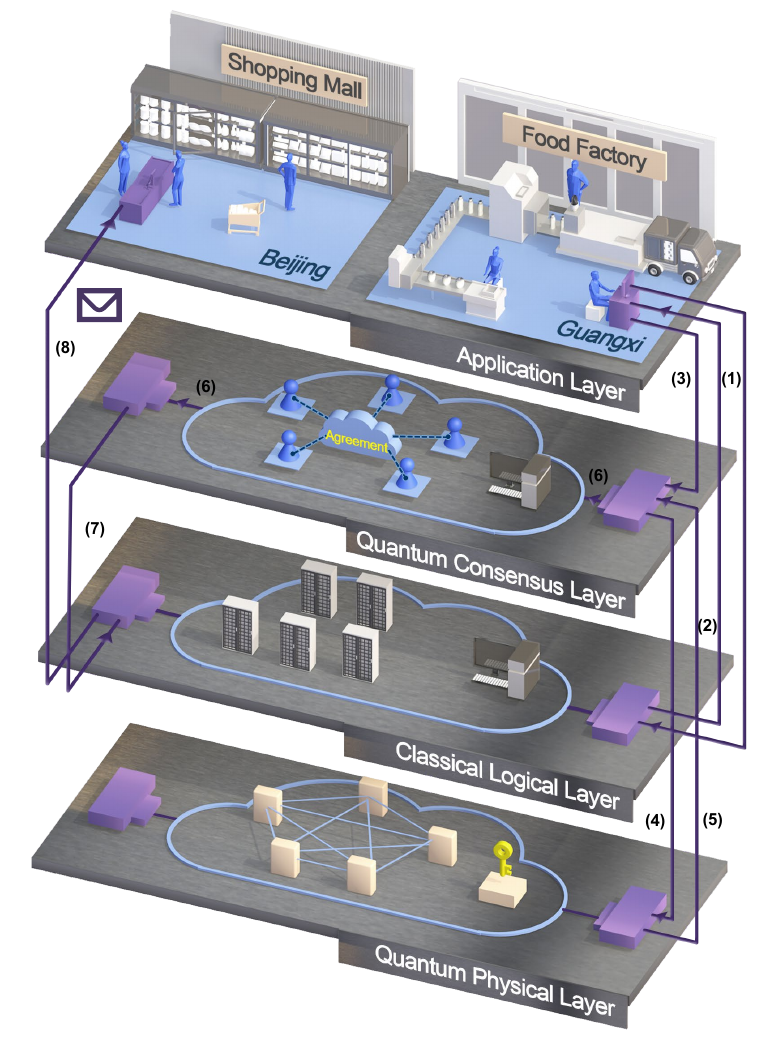}
\caption{\textbf{ Illustration of quantum blockchain architecture. } It consists of four layers: the Application Layer, Quantum Consensus Layer, Classical Logical Layer, and Quantum Physical Layer. The end-to-end workflow proceeds as follows:  the client sends a transaction proposal to peer nodes in the classical logical layer (1), which verify the syntax and semantics, execute the endorsement chaincode, and return signed endorsements (2). Once sufficient endorsements have been collected, the client submits the verified transaction to ordering nodes in the quantum consensus layer (3), which check local key resources, if found insufficient, request fresh quantum key material from the quantum physical layer (4--5), and then execute the quantum consensus protocol to reach a distributed agreement on the global order of the transaction batch (6). The ordered transactions are packaged into a block and sent back to the classical logical layer (7). Finally, the peer nodes validate the block, commit it to the ledger, and return the execution results to the client (8), thereby completing the transaction lifecycle.}
	\label{fig:architecture}
\end{figure}

The proposed four-layer quantum blockchain architecture is illustrated in Fig.~\ref{fig:architecture}. It consists of four layers: the application, quantum consensus, classical logical, and quantum physical layers. 

The quantum physical layer constitutes the foundational infrastructure of the quantum blockchain platform. Its primary function is to supply information-theoretically secure key resources to user nodes by implementing practical quantum cryptographic protocols over a deployable quantum communication network. To enable stable and scalable key generation with current technology, each user node is equipped with a quantum state transmitter and receiver, deployed over commercial telecommunication fiber links. Building on this hardware foundation, a state-of-the-art quantum cryptographic KGP~\cite{2020-Xu,2020-Pirandola}, such as BB84~\cite{1984-BB84,2023-Li-QKD}, measurement device independent~\cite{2012-Lo,xie2022breaking,zeng2022mode} and twin-field~\cite{2018-Lucamarini,2018-Wang-TF,2018-Ma-PM,cui2019twin,2019-Curty} protocols, is employed to implement the core functionality of this layer.

The quantum consensus layer is built upon the information-theoretically secure key resources generated by the quantum physical layer and is responsible for establishing agreement among independent and mutually untrusted nodes on the ordering and content of messages required by the application layer. Depending on the functional requirements of the blockchain application, the architecture supports the deployment of suitable quantum consensus protocols. QBA is one of the most commonly used quantum consensus protocols for distributed systems, in which each node operates independently and possesses equal capabilities, enabling these mutually untrusted nodes to reach consensus on messages in the network. In general, a QBA protocol consists of two main phases. In the first phase, a designated main node distributes its proposed message to the replica nodes. In the second phase, each replica node derives the main node's   initial message based on the information received during the first phase. By exploiting quantum resources in these two phases, such as multipartite quantum correlations~\cite{2001-Fitzi,2008-Gaertner} and quantum digital signatures (QDS)~\cite{2023-Weng,2024-Jing}, QBA protocols relax the classical independence assumptions on pairwise communication channels and thereby achieve enhanced fault tolerance and security compared with their classical counterparts.

The application/classical logical layer transforms the information-theoretically secure and highly fault-tolerant agreement established by the quantum consensus layer into practical, real-world services. In doing so, the application initiates workflows based on client requests, while the classical logical layer ensures transaction verification, endorsement, and ledger state updates in accordance with Hyperledger Fabric's transaction lifecycle.

\begin{figure*}[t!]
	\begin{center}
		\includegraphics[width=\textwidth]{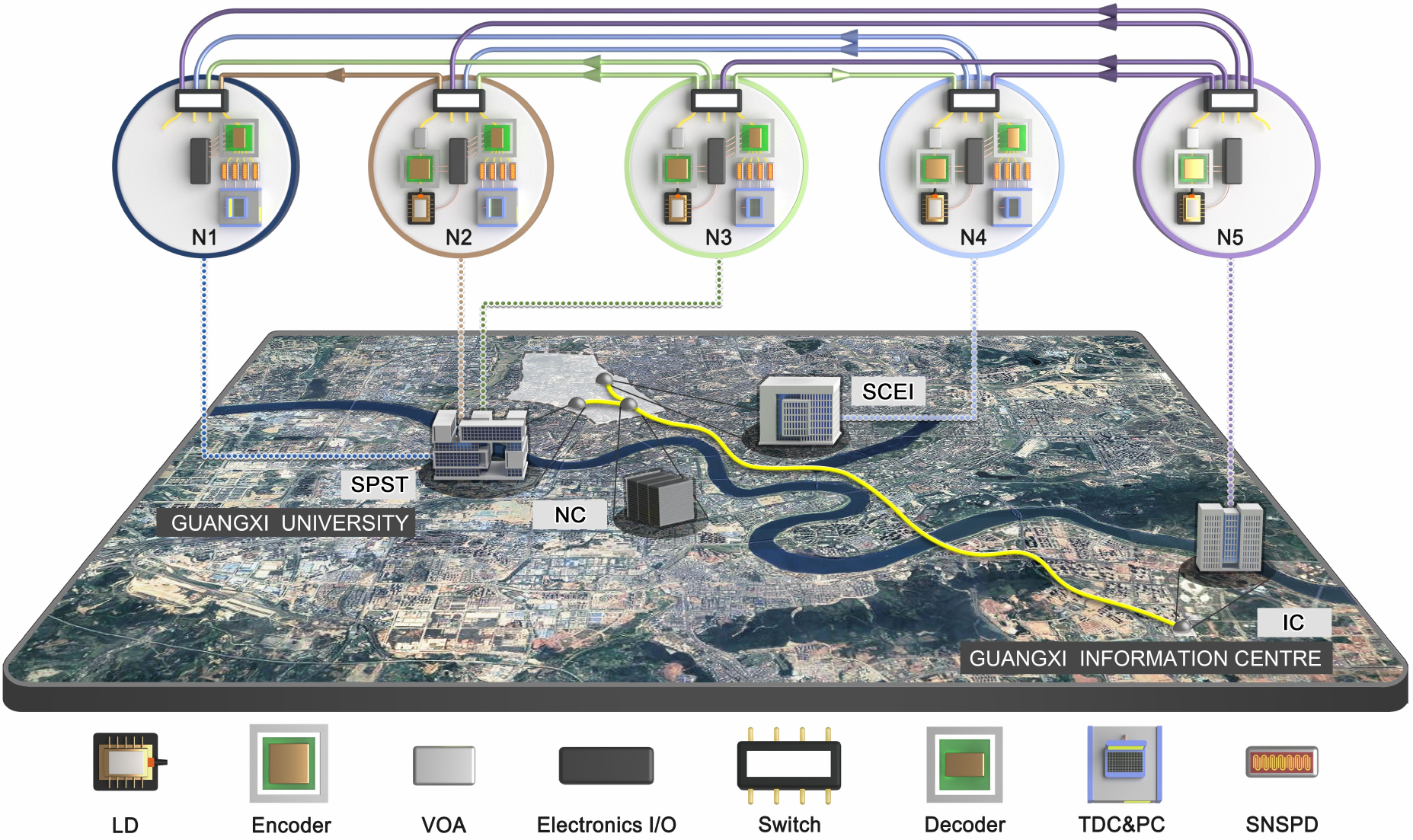}
	\end{center}
	\caption{\textbf{Aerial view of the intercity quantum blockchain.} \textbf{Bottom:} The experimental network consists of five nodes deployed across Nanning in the Guangxi Zhuang Autonomous Region, China. Nodes~N1--N3 are located in three separate laboratories within the School of Physical Science and Technology (SPST) at Guangxi University; Node~N4 is located in the School of Computer, Electronics and Information (SCEI); and Node~N5 is located at the Guangxi Zhuang Autonomous Region Information Center (IC). All five nodes connect to the Network Center (NC) of Guangxi University through dedicated fiber links, forming a star-like network topology. \textbf{Top:} Building on this network topology, the five nodes form a fully connected logical configuration, resulting in ten point-to-point communication paths. To clearly illustrate this logical interconnection in the figure, an optical switch (Switch) is included as a schematic representation, although it is not part of the actual deployed system. Regarding device configuration, Node~N1 is equipped with an integrated quantum signal receiver; Node~N5 is equipped with an integrated quantum signal transmitter; and Nodes~N2--N4 are each equipped with both an integrated transmitter and an integrated receiver. The integrated quantum signal transmitter consists of a laser diode (LD), an encoder chip (Encoder), a variable optical attenuator (VOA), and the corresponding electronic driver (Electronics~I/O). The integrated quantum signal receiver comprises a polarization-tracking decoder chip, a multi-channel superconducting nanowire single-photon detector (SNSPD), a time-to-digital converter (TDC), a personal computer (PC), and the associated electronic driver. Map data: Google Earth.  }
	\label{fig:network}
\end{figure*}

\section{Network deployment}

To demonstrate the feasibility of the constructed quantum blockchain, we deploy the quantum physical layer over the metropolitan fiber network in Nanning, Guangxi Zhuang Autonomous Region, China (see Fig. \ref{fig:network}).  The network has five nodes and adopts a fully connected topology, enabling any pair of nodes to share secure key bits and exchange authenticated messages in the quantum consensus layer with unforgeability and non-repudiation using one-time universal$_{2}$ hashing quantum digital signatures (OTUH-QDS)~\cite{2023-Yin}.

Nodes N1, N2, and N3 are located in three separate laboratories within the School of Physical Science and Technology (SPST) at Guangxi University. The average separation between these nodes is 0.5~km, corresponding to an average channel loss of 1.0~dB. Node N4 is located in the School of Computer and Electronic Information (SCEI) at Guangxi University, while Node N5 is housed in the Guangxi Zhuang Autonomous Region Information Center (IC). The point-to-point distances from SPST to SCEI and IC are 2.1~km and 30.7~km, respectively.

The five nodes (N1-N5) connect to a central switch in the Network Center (NC) of Guangxi University via dedicated fiber links, forming a star-like topology (see the bottom of Fig.~\ref{fig:network}). This architecture effectively emulates a fully connected configuration, enabling all ten unique node pairs to establish point-to-point channels (see the top of Fig.~\ref{fig:network}). The ten links have an average length of 13.1~km and an average loss of 13.2~dB.

To distribute information-theoretically secure key bits
between any pair of nodes with minimal equipment, each
node is equipped with different types of integrated quantum devices, as shown in the top of Fig.~\ref{fig:network}.

Node~N1 is equipped with an integrated quantum signal receiver, while Node~N5 is equipped with an integrated quantum signal transmitter. Nodes N2, N3, and N4 require both state preparation and measurement and are therefore equipped with both an integrated quantum signal transmitter and receiver.

The integrated quantum  transmitter comprises a phase-randomized pulsed laser, an encoding chip with decoy-state intensity and polarization modulation, an off-chip variable optical attenuator, and driving circuits for both the laser and the chip. The integrated quantum  receiver includes a polarization-decoding chip with polarization-tracking capability, superconducting nanowire single-photon detectors (SNSPDs), a time-to-digital converter (TDC), a personal computer (PC), and the associated driving circuits. Both the encoder and decoder chips are designed and fabricated on a silicon-on-insulator platform and integrate thermo-optic and electro-optic tunable components that enable high-speed preparation and measurement of polarization-encoded quantum states. Further details for the point-to-point setup are shown in Fig.~\ref{fig-chip} of the Appendix~\ref{app:KGP}, and those for the experimental implementation are provided in Appendix~\ref{app:KGP}.

\section{Recursive QBA protocol}

We adopt a recursive QBA protocol~\cite{2023-Weng} which employs OTUH-QDS in the quantum consensus layer, thereby surpassing the classical one-third fault-tolerance bound and achieving fault tolerance approaching one-half.

We summarize a five-party implementation ($N=5$, including a main node and four replica nodes) used in our experimental demonstration, which tolerates up to two malicious participants ($f=2$). Detailed descriptions of the OTUH-QDS and the QBA protocols are provided in Appendix~\ref{app:OTUHQDS} and Appendix~\ref{app:QBA}, which include a step-by-step example for the five-party and general $N$-party settings, as well as an analysis of fault tolerance, communication complexity, and consensus rate.

The five-party protocol proceeds recursively and comprises two main phases: a \textit{Broadcasting Phase} for information exchange and a \textit{Gathering Phase} for decision-making. The core operation of this protocol is the multicast round, in which a designated primary distributes a message to a specific group of backup nodes using OTUH-QDS~\cite{2023-Yin,2025-Du}. In a multicast round, one backup node is designated as the forwarder, while each remaining backup node serves as the verifier one by one. The primary (signer), the forwarder, and the verifiers jointly execute the OTUH-QDS process to securely transmit and authenticate the message. This procedure repeats until each backup node has acted as the forwarder once. Owing to the unforgeability and non-repudiation properties of QDS~\cite{bozzio2025quantum}, the multicast round guarantees that transmitted messages cannot be forged or repudiated. 

The Broadcasting Phase features a recursive structure of depth $2$, which is equal to the maximal number of malicious nodes the system can tolerate. At depth $d=1$, the main node acts as the primary node, and four replica nodes are the backup nodes. The main node distributes a message to all replica nodes via a multicast round. At depth $d=2$, each replica node acts as a new primary in a multicast round, with the other three replica nodes being the backup nodes, to announce the message it received from the main node at depth $d=1$ to the other replica nodes.

The goal of the Gathering Phase is for each replica node to independently derive the same final output from the collection of messages announced during the Broadcasting Phase. Decisions are reached through a bottom-up majority-voting process. Each replica node first applies a majority vote to each of its bottom message lists recorded at the deepest layer ($d=2$), thereby forming a final message list at depth $d=1$. The resulting values represent the replica node's reconstruction of the messages originally received by other replicas from the main node. Finally, each replica node applies a majority vote to its final message list at depth $d=1$ to obtain the consensus result. This recursive majority-voting procedure guarantees that, even if a malicious main node $N_1$ distributes conflicting messages, all honest replicas independently reconstruct identical message lists of depth $d=1$ and give the same final outcome.

\section{Food traceability system}

To achieve meaningful application throughput for real-world tasks, we develop a quantum-blockchain-based food traceability system that is broadly applicable to global supply chain management. The system enables transparent, secure, and efficient farm-to-table tracking, even in the presence of adversaries equipped with quantum computational capabilities.

The system adopts a microservice architecture, with the backend developed using the Spring Boot framework. It is  integrated with the Hyperledger Fabric Java SDK to enable seamless interaction with the quantum blockchain network. All backend operations run within a Java environment. The frontend, built with Vue3 and HBuilderX, provides an intuitive and efficient user interface that facilitates smooth interaction between users and the traceability platform. The comprehensive architectural design is presented in Appendix~\ref{app:FTS} and Fig.~\ref{fig:Interaction_architecture} of the Appendix, the source code is publicly available at GitHub repository \url{(https://github.com/zmx-creator/Quantum-Food-traceability-system)}, and operation of the software is illustrated in audiovisual tutorials. A schematic overview of the system is presented below.

As illustrated in Fig.~\ref{fig:application}, the traceability process begins at the user interface, where users input food data (1), including product details, ingredient origins, logistics information, and other supply chain attributes. The submitted information is then sent to the backend service (2), which validates the data and invokes the appropriate chaincode functions to write product, ingredient, and logistics records or to query existing transactions. Before initiating consensus procedures, the system performs data normalization, format conversion, and preliminary verification to ensure data accuracy and completeness. Once the data is prepared and verified, the ordering nodes execute the QBA quantum consensus algorithm to reach a distributed agreement on the total order of the transaction batch (3). This step ensures that each ledger update is consistent across nodes and resistant to tampering.

Following successful consensus, the finalized transaction batch is encapsulated into a block and committed to the blockchain (4), becoming part of a permanent, immutable ledger that serves as the authoritative source of truth for all traceability records. Authorized users can subsequently query traceability results by issuing search requests to the ledger (5), retrieving comprehensive supply chain histories that link each product to its origin and handling records. In the View Results stage (6), the system presents the retrieved information through a user-friendly interface, providing clear product details, complete traceability timelines, and visual indicators that enhance transparency and consumer confidence.

This end-to-end system, spanning data acquisition to result visualization, is underpinned by a secure and highly fault-tolerant quantum blockchain architecture, providing a robust and transparent framework that ensures supply chain integrity.

\begin{figure}[t!]
	\begin{center}
		\includegraphics[width=0.5\textwidth]{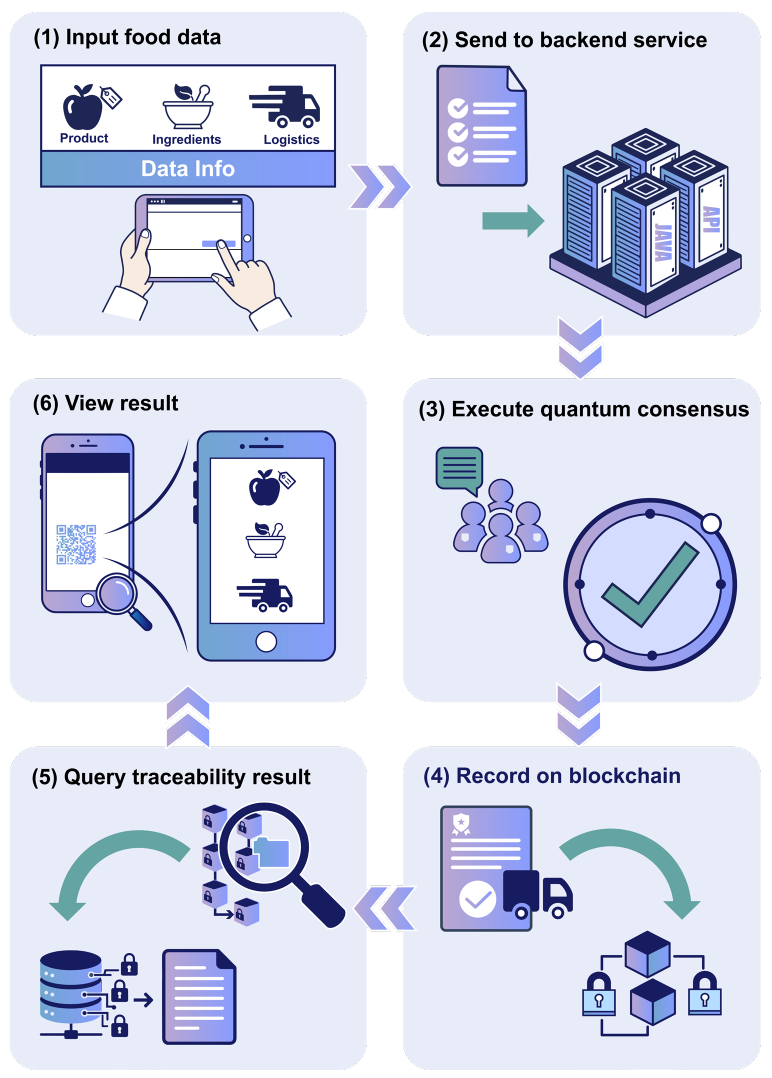}
	\end{center}
	\caption{\textbf{ Schematic diagram of the food traceability system. } The system operates through six stages: (1) Input food data, where product, ingredient, logistics, and related information are provided by users; (2) Send to backend service, in which the data is transmitted for structured processing; (3) Execute quantum consensus, where participating nodes, via the proposed QBA algorithm, reach agreement on the total order of transactions to ensure consistent and tamper-proof ledger updates; (4) Record on blockchain, permanently storing confirmed data in an immutable ledger; (5) Query traceability result, allowing stakeholders to retrieve supply chain records; and (6) View result, presenting clear product details and complete traceability histories through user-friendly interfaces. }
	\label{fig:application}
\end{figure}

\begin{figure}[t!]
	\begin{center}
		\includegraphics[width=0.5\textwidth]{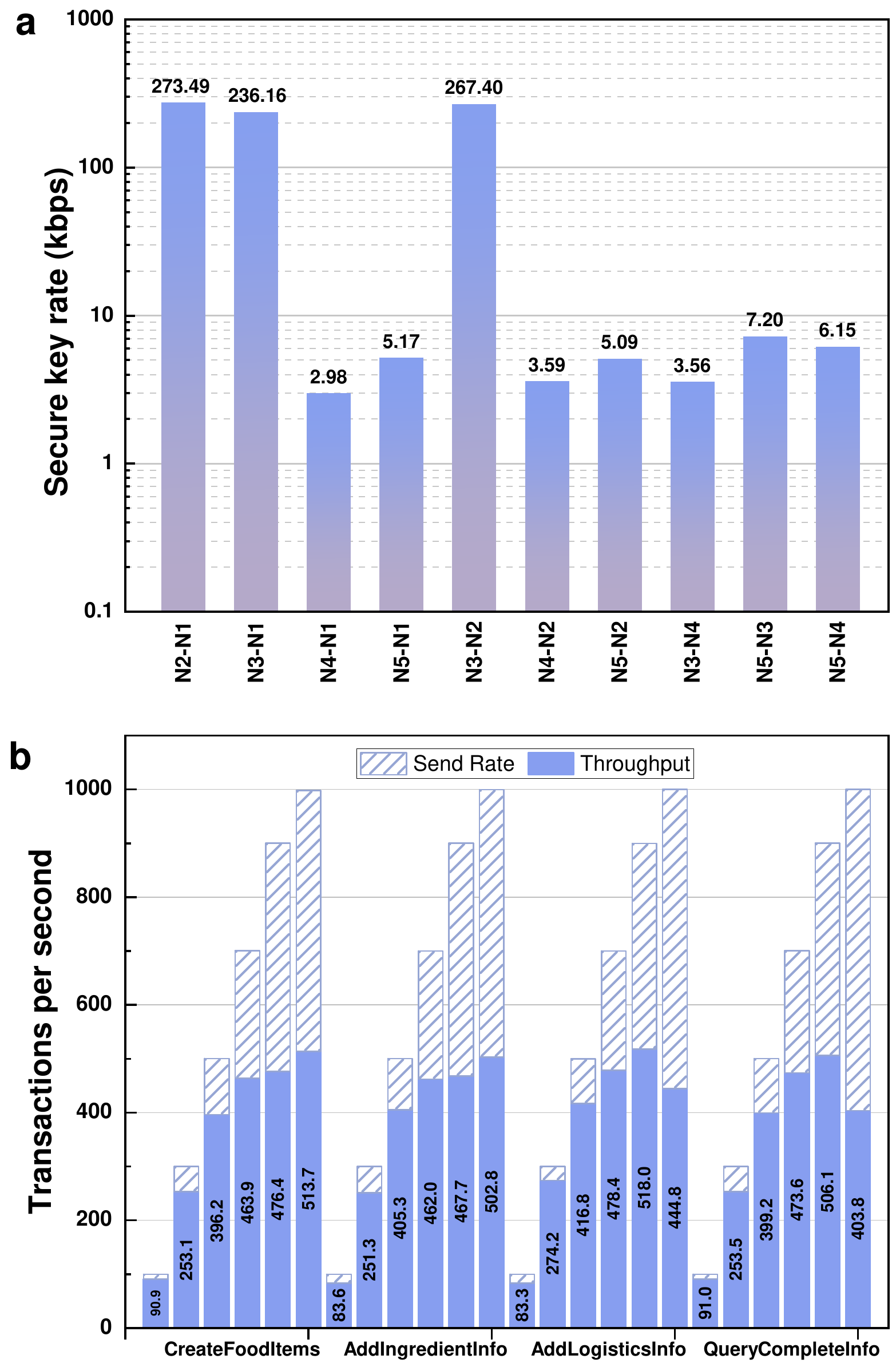}
	\end{center}
	\caption{\textbf{Performance of the hybrid quantum blockchain.} \textbf{a.} The bar chart illustrates the secure key rates across the ten point-to-point links within the network. The horizontal axis represents the node pairs, while the vertical axis corresponds to the secure key rates. \textbf{b.} The stacked bar chart showing the send rate (upper segment) and throughput (lower segment) in transactions per second for four transaction types in the food traceability system. Metrics were obtained using the Hyperledger Caliper benchmarking tool under varying load conditions (1,000-10,000 transactions). The send rate represents the rate at which transactions were issued to the network, while throughput denotes successfully committed transactions. Differences between send rate and throughput reflect resource contention under high load.} 
	\label{fig:result-QKD}
\end{figure}

\section{Experimental results}

We experimentally evaluate the system-level performance of a hybrid quantum blockchain implemented with integrated photonic circuits. We first estimate the secure key rate of the quantum physical layer across ten point-to-point links deployed in a metropolitan-area network, where each user pair distributes information-theoretically secure keys using a one-decoy-state BB84 protocol with globally optimized parameters. With collected 10 Mega valid detection events in the \(Z\)-basis  and finite-key analysis, as shown in Fig.~\ref{fig:result-QKD}a, all links achieve secure key generation after error correction and privacy amplification, yielding secure key rates $R$ exceeding 2.98 kbps for all links and reaching up to 273.49 kbps. Further detailed experimental parameters for each link are provided in Appendix~\ref{app:TS1}.

Leveraging these secure keys, we benchmark the quantum consensus layer executing a recursive QBA protocol using OTUH-QDS. By exploiting the recursive QBA construction, the five-user quantum blockchain tolerates up to two faulty parties in a five-party network, satisfying the fault-tolerance condition $N \ge 2f+1$. The asymmetric relationship of three-party OTUH-QDS employed in the recursive QBA removes the independence of point-to-point communication in the network, thus enabling the protocol to surpass the classical one-third fault-tolerance limit~\cite{2023-Weng}.

We further benchmark the consensus rate ($CR$) of the five-node quantum consensus layer for agreement on a 1-Mbit message. The $CR$, defined as the number of consensus instances per second, is fundamentally limited by the minimum secure key rate $R$ across the network and the protocol's communication complexity $C$~\cite{2023-Weng},
\[
CR=\min_E(R)/(3nC).
\]
Here, $E$ denotes the set of all point-to-point links and $\min_E(R)$ means the minimal key rate across all links. The communication complexity $C$ denotes the number of required QDS, and $n$ is the length of the signature ($3n$-bit secret key is needed for an $n$-bit signature~\cite{2023-Yin}). For the five-party demonstration, $C=36$ and $n=64$, corresponding to a consensus-layer security of $\sim10^{-11.4}$; with a bottleneck key rate of 2.98 kbps, the resulting consensus rate is approximately 0.43 transactions per second.

Building on the quantum layers, we evaluate a deployed food traceability application using Hyperledger Fabric~1.4.4 and Hyperledger Caliper~\cite{2023-Al-Sumaidaee,2021-Guo}, running on an Ubuntu~20.04 virtual machine with a dual-core processor and 16~GB of RAM. The results are summarized in Fig.~\ref{fig:result-QKD}b, with detailed data provided in Appendix~\ref{app:TS2}. End-to-end benchmarking shows zero failed transactions under workloads of up to 10{,}000 operations. The system achieves a peak throughput of nearly 500\,transactions per second, with latency increasing from 0.9\,s to 8.7\,s under heavy load, validating the practicality of the proposed hybrid quantum blockchain.


\section{ Conclusion and discussion}
We experimentally demonstrate a hybrid quantum blockchain integrating a photonic quantum physical layer with a quantum consensus layer. Leveraging information-theoretically secure quantum protocols and a recursive QBA, the system achieves fault tolerance near one-half, surpassing the one-third limit of classical blockchains. This is the first implementation of a hybrid quantum blockchain deployable over commercial classical telecommunication infrastructure, demonstrating a practical route toward secure and fault-tolerant distributed ledgers.

Our five-node network achieves secure key rates up to 273.49~kbps and consensus on 1-Mbit messages at 0.43~tps, validating multi-node agreement under realistic conditions. The food traceability application further illustrates the platform's utility, with end-to-end data acquisition, block recording, and querying at a throughput of 500~transactions per second without failed transactions, highlighting potential for high-volume, distributed systems.

This architecture lays a foundation for diverse applications, from supply chain management to digital assets and the Metaverse, providing a scalable framework for quantum-enhanced distributed systems.

\section{ Acknowledgements}
We thank Pingping Li for drawing the figures. This study was supported by Quantum Science and Technology-National Science and Technology Major Project (Grant No. 2025ZD0300200), the National Natural Science Foundation of China (Grants No.~12522419, No.~U25D8016, No.~12274223, No.~62171144, No.~62031024, and No.~62171485), Guangxi Science Foundation (Nos. 2025GXNSFAA069137, 2026GXNSFHA00640303), Guangdong Basic and Applied Basic Research Foundation (2024B1515120030), and Bagui Scholars Programme (W.X.-G., GXR-6BG2424001, GXR-1BGQ2424005).

\appendix
\setcounter{figure}{0}
\setcounter{table}{0}
\renewcommand{\thefigure}{S\arabic{figure}}
\renewcommand{\thetable}{S\arabic{table}}
\renewcommand{\theHfigure}{S\arabic{figure}}
\renewcommand{\theHtable}{S\arabic{table}}

\section{ KGP implementation}\label{app:KGP}

\begin{figure*}[t!]
	\begin{center}
		\includegraphics[width=\textwidth]{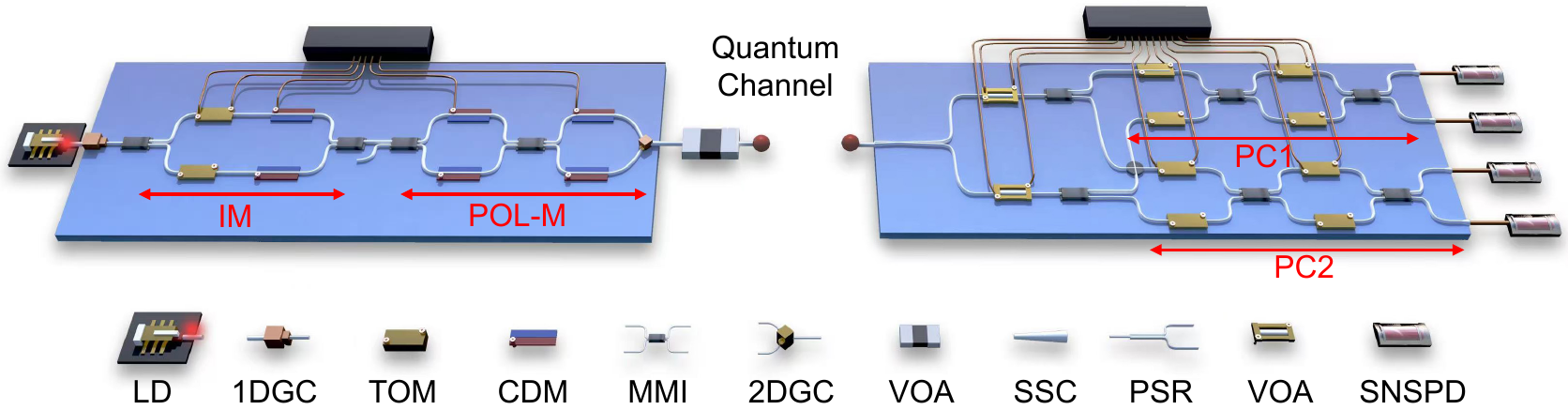}
	\end{center}
	\caption{\textbf{ Integrated photonic circuits for the encoding and decoding of polarization-encoded quantum states.} The integrated quantum signal transmitter (left) encodes quantum states onto weak coherent optical pulses and transmits them through the quantum channel to the integrated quantum signal receiver (right), which performs quantum state decoding and detection. The transmitter consists of a laser diode (LD), a silicon-based encoder chip, and an off-chip variable optical attenuator (VOA). The silicon encoder chip integrates one-dimensional and two-dimensional grating couplers (1DGC and 2DGC) for optical input and output, as well as an intensity modulator (IM) and a polarization modulator (POL-M) constructed from multimode interferometers (MMIs), thermo-optic modulators (TOMs), and carrier-depletion modulators (CDMs), which enable decoy-state intensity modulation and polarization-state preparation. The receiver comprises a silicon-based decoder chip with polarization-tracking capability and a multi-channel superconducting nanowire single-photon detector (SNSPD). The silicon decoder chip incorporates spot-size converters (SSCs) for optical coupling, a polarization splitter-rotator (PSR) for mapping polarization states to path states, a VOA for compensating polarization-dependent loss, and two polarization controllers (PC1 and PC2) built from MMIs and TOMs to provide polarization calibration and stable decoding. }
	\label{fig-chip}
\end{figure*}

The quantum physical layer deploys integrated quantum  transmitters and receivers at user nodes, as illustrated in Fig.~\ref{fig-chip}. Leveraging these integrated devices, the one-decoy-state BB84 key generation protocol (KGP)~\cite{2018-Rusca} is implemented over commercial fiber networks to distribute information-theoretically secure quantum keys to user nodes within the quantum blockchain.  

\subsection{Setup}

At the transmitter, a laser diode (LD) generates phase-randomized optical pulses, which are coupled into the Si photonic encoder chip via a 1-dimensional grating coupler (1DGC). The encoder chip integrates both an intensity modulator (IM) and a polarization modulator (POL-M). The IM is implemented using a Mach-Zehnder interferometer (MZI), which incorporates two multimode interferometers (MMIs) and two types of phase modulators: a pair of thermo-optic modulators (TOMs) for setting static phase bias, and a pair of carrier-depletion modulators (CDMs) for high-speed phase modulation. The IM enables random modulation between signal states and decoy states.

The output of the IM is connected through a waveguide to the POL-M, which performs polarization-state modulation. The POL-M consists of a MZI driven by a pair of CDMs, which connects to another pair of external CDMs to implement path encoding. The path-encoded information is subsequently transformed into polarization encoding by a two-dimensional grating coupler (2DGC) and coupled into the external fiber. The POL-M generates the four BB84 polarization states $\left| \psi  \right\rangle  =( \left| H \right\rangle  + {e^{i\theta }}\left| V \right\rangle)/\sqrt{2} ,\theta  \in \{ 0,{\pi  \mathord{\left/
		{\vphantom {\pi  2}} \right.
		\kern-\nulldelimiterspace} 2},\pi ,{{3\pi } \mathord{\left/
		{\vphantom {{3\pi } 2}} \right.
		\kern-\nulldelimiterspace} 2}\}$,    where $\theta\in\{0, \pi\}$ ($\theta\in\{\pi/2, 3\pi/2\}$)  corresponds to the $Z$ ($X$) basis. The signal pulses encoded by the chip are attenuated to the single-photon level through an off-chip variable optical attenuator (VOA) and subsequently transmitted to the receiver.

At the receiver, the incoming optical pulses are coupled from the fiber into a Si photonic decoder chip with polarization tracking via a spot-size converter (SSC). The polarization-encoded information carried by the pulses is transformed into on-chip path encoding using a polarization splitter-rotator (PSR). The encoded signals are passively distributed to the \(Z\)- or \(X\)-basis measurement paths via two symmetric MMIs. The \(Z\)- and \(X\)-basis measurements are implemented by two polarization controllers (PC1 and PC2), each of which comprises a pair of TOMs bridging a MZI, driven by an additional pair of TOMs. By precisely tuning the drive voltages of these TOMs, projective measurements in the \(Z\) and \(X\) bases are performed on-chip. The measured optical pulses are subsequently routed via the SSC and coupled to superconducting nanowire single-photon detectors (SNSPDs) for detection.

Both the encoder and decoder chips are designed and fabricated using standard silicon photonics processes provided by a commercial foundry, with robust packaging ensuring long-term operation. The encoder chip has dimensions of \(6 \times 3~\mathrm{mm^2}\) and is packaged in a butterfly configuration, with a total volume of \(20 \times 11 \times 5~\mathrm{mm^3}\). The decoder chip has a footprint of \(1.6 \times 1.7~\mathrm{mm^2}\) and is packaged in an on-board assembly, with a total volume of \(3.95 \times 2.19 \times 0.90~\mathrm{cm^3}\). The CDMs in the encoder are driven by an arbitrary waveform generator, facilitating high-speed quantum state encoding. All TOMs in both the encoder and decoder chips are precisely controlled by programmable linear DC sources.

\subsection{Device characterization}
We conduct a detailed characterization of the encoding and decoding chips comprising the integrated quantum signal transmitters and receivers deployed in the quantum physical layer. The characterization is performed using optical pulses with a central wavelength of 1549.17 nm, a pulse width of 200 ps, and a repetition rate of 50 MHz. The test results show that the dynamic extinction ratio of the IM is approximately 18 dB, which satisfies the requirements of the 1-decoy-state BB84 KPG scheme. Using the encoder chip, four polarization states corresponding to the $Z$ basis $\{|H\rangle, |V\rangle\}$ and the $X$ basis $\{|D\rangle, |A\rangle\}$ are generated, and subsequent measurements using the decoder chip yield an average polarization extinction ratio of approximately 23 dB. This demonstrates that the encoder-decoder subsystem effectively supports low-error-rate operation of KGP systems across all links. The total insertion loss of the decoder chip is measured to be approximately 7.9 dB, and the basis-selection probabilities for the $Z$ basis are  close to 50\%. The detection efficiency of the SNSPDs is uniformly calibrated to 60\%. 

\subsection{Clock synchronization}
We employ a qubit-based frame synchronization method~\cite{2025-Chen} to achieve clock synchronization between network nodes. This scheme directly utilizes the polarization states required by the BB84 KGP, which are prepared and measured on-chip, thereby enabling inter-node clock synchronization without additional hardware. The procedure is detailed as follows.

At the transmitter, a synchronization string $s^{A}$ of length $L$ exhibiting periodic correlation properties is constructed. The synchronization string is divided into $L$ individual synchronization qubits, each followed by $M$ randomly encoded qubits, forming a basic data block of length $M+1$. Concatenating all $L$ data blocks sequentially yields a synchronization frame with total length $N_f = (M+1)L$. The synchronization string $s^{A}$ is publicly announced to the receiver via an authenticated channel and can be reused, whereas the random qubit sequences are updated frame by frame for key generation.

Subsequently, the integrated quantum signal transmitter encodes the qubit sequence according to the local optical pulse period $\tau^{A}$. Each synchronization qubit takes a value of $+1$ or $-1$ and is encoded at the signal-state intensity in the $Z$ basis. Random qubits are encoded according to the decoy-state BB84 protocol, with intensity randomly chosen between signal and decoy states and the basis randomly chosen between the $Z$ and $X$ bases. The encoded quantum states are transmitted to the receiver through the quantum channel.

Upon arrival at the receiver, quantum pulses are measured by the integrated quantum receiver, and the arrival times of all detection events are recorded. The receiver first samples the detection events and applies a fast Fourier transform to obtain an initial estimate of the transmitter clock period, denoted $\tau_{0}^{B}$. This estimate is refined using the least-trimmed-squares method, yielding a more accurate estimate of the transmitter clock period $\tau_{B}$. The recovered clock period $\tau_{B}$ maps the continuous detection timestamps onto discrete qubit indices and serves as the periodic basis for the subsequent time-offset estimation.

 In the time-offset estimation stage, the receiver retains only the detection outcomes obtained in the $Z$ basis, assigning a value of $+1$ to detections corresponding to the $\lvert H\rangle$ state and a value of $-1$ to those corresponding to the $\lvert V\rangle$ state. Detection events in the $X$ basis, as well as undetected events, are assigned a value of $0$. According to these rules, the detection events are mapped to a detection string $s^{B}$ of length $N_f$. According to the synchronization frame structure, the detection string $s^{B}$ is partitioned into $L$ data blocks of length $M+1$ and rearranged into a $(M+1)\times L$ matrix $\mathbf{S}$. For each row $s_i^{B}$ ($i=1,2,\ldots,M+1$) of the matrix $\mathbf{S}$, the cross-correlation with the public synchronization string $s^{A}$ is computed. From the position of the optimal correlation peak, the locations of the synchronization qubits in the received sequence are identified, thereby recovering the absolute time offset between the transmitter and the receiver and enabling clock synchronization between remote nodes.

\subsection{1-decoy state BB84 KGP}
In this work, we implement the 1-decoy-state BB84 KGP~\cite{2018-Rusca} to distribute information-theoretically secure quantum keys using integrated quantum devices deployed over a commercial fiber network. Here, we illustrate the key-generation process using a pair of user nodes:

\textit{1. Transmission.}
The transmitter randomly selects a bit value and a basis \(\lambda \in \{Z, X\}\) according to predefined probabilities \(P_\lambda\). For each phase-randomized pulse, an intensity \(k \in \{\mu, \nu\}\) (signal or decoy) is chosen with probability \(P_k\), and the qubit is encoded into the corresponding quantum state \(\{|0_Z\rangle, |1_Z\rangle, |0_X\rangle, |1_X\rangle\}\). Each pulse is attenuated to the single-photon level and transmitted to the receiver via the quantum channel. 

\textit{2. Detection.}
Upon receiving the quantum signals, the receiver independently and randomly selects a measurement basis \(\lambda \in \{Z, X\}\) according to predefined probabilities and records the resulting detection events. For each valid detection, the receiver records the employed measurement basis and the corresponding bit value.

\textit{3. Basis reconciliation.}
 The basis choices are publicly disclosed over an authenticated classical channel, and only detection events corresponding to matching bases are retained. This process yields the counts \(n_{\lambda,k}\) for each measurement basis \(\lambda\) and intensity setting \(k\).

\textit{4. Parameter estimation.}
The two parties publicly disclose the bit values of detection events measured in the \(X\) basis to evaluate the corresponding error counts \(m_{X,k}\). Using the 1-decoy-state method~\cite{2018-Rusca}, they estimate the lower bounds on the numbers of vacuum and single-photon events, denoted by \(s_{Z,0}^{l}\) and \(s_{Z,1}^{l}\), respectively, as well as the upper bound on the single-photon phase error rate \(\phi_{Z}^{u}\) associated with the \(Z\) basis. 

\textit{5. Error correction and verification.} 
The two parties use the bit information obtained in the $Z$ basis to generate the raw key. They disclose $\lambda_{\rm EC}$ bits of information to perform an error correction procedure capable of correcting errors at the expected quantum bit error rate (QBER) $E_Z$. To ensure that the sender and receiver hold identical keys with correctness parameter $\varepsilon_{\rm cor}$, they perform an error verification step using two universal hash functions, which disclose $\left \lceil \log_{2}{1/\varepsilon_{\rm cor}} \right \rceil$ bits of information.

\textit{6. Privacy amplification.} 
 The communicating parties further estimate the number of bits to be sacrificed based on finite-key analysis. They then apply a universal hash function to the error-corrected string, ultimately generating a secure key of length $l_{\rm sec}$.

\subsection{Secret key rate calculation}
To maximize the overall secure key rate, all implementation parameters are globally optimized. Using the optimal parameters, we employ a periodic correlation code of length \(5 \times 10^{4}\) with a ratio of \(M = 9{:}1\), where every ten qubits contain nine signal qubits and one synchronization qubit for system configuration. For each link, we collect \(10^{7}\) \(Z\)-basis detection events as the raw key, and then perform error correction to ensure key consistency between all pairs of user nodes. We subsequently conduct finite-key analysis following the method in Ref.~\cite{2018-Rusca}.
\begin{equation}\label{R_finite}
	\begin{split}
		R \le \Big\{ &s_{z}^{(L,0)} + s_{z}^{(L,1)} \bigl[ 1 - h(\phi_{z}^{U}) \bigr]
		- \mathrm{leak}_{\mathrm{EC}} \\
		&-\, 6\log_{2}\!\bigl(19/\varepsilon_{\mathrm{sec}}\bigr)
		- \log_{2}\!\bigl(2/\varepsilon_{\mathrm{cor}}\bigr) \Big\} q/ t,
	\end{split}
\end{equation}
here, $t$ represents the duration of a single data-acquisition period; \(s_{z}^{(L,0)}\) and \(s_{z}^{(L,1)}\) denote the lower bounds on the effective vacuum and single-photon event contributions measured in the \(Z\) basis; $\phi_{z}^{U}$ represents the upper bound of the phase-error rate in the $Z$ basis; $\mathrm{leak}_{\mathrm{EC}}$ denotes the amount of information disclosed during the error-correction process; $\varepsilon_{\mathrm{sec}}$ and $\varepsilon_{\mathrm{cor}}$ denote the secrecy and correctness parameters, respectively; and $h(x)$ denotes the binary Shannon entropy function; $q = M/(M+1)$ denotes the fraction of qubits used for  key generation.

\subsection{Error correction}
After obtaining the raw keys, the two parties perform error correction on key blocks of length \(10^{7}\) to generate identical keys. In the first round, each block is processed sequentially and divided into sub-blocks according to the bit error rate \(E_Z\). The sender computes the parity of each sub-block and transmits it to the receiver via an authenticated channel. The receiver compares the received parity with its local calculation.  If a match is confirmed, they proceed; otherwise, receiver signals the discrepancy, and they cooperatively perform a binary search over the authenticated public channel to localize and correct the erroneous bit. In subsequent rounds, the keys are randomly permuted using a shared random sequence, and the sub-block size is doubled. Parity information is exchanged over an authenticated public channel, and any detected discrepancies are corrected through a binary search to identify erroneous bits. All publicly disclosed parity information accumulated during the procedure is used to quantify the total error-correction leakage, denoted by \(\mathrm{leak}_{\mathrm{EC}}\). Iterations continue, with the key indices reshuffled and the block size adjusted after each round, until convergence to identical raw keys is achieved.

\subsection{Privacy amplification}
Based on the secure key rate determined via finite-key analysis, the sender and receiver perform privacy amplification on the error-corrected keys to generate the final secure key. The sender constructs a Toeplitz matrix of size \(n \times l\), where the ratio \(n/l\) corresponds to \(10^7\) divided by the secure key length \(L_{\mathrm{sec}}\) (\(L_{\mathrm{sec}} = R \cdot t\)). This Toeplitz matrix is generated from \(n+l-1\) random bits transmitted over an authenticated channel. Both parties then multiply their respective error-corrected keys by the Toeplitz matrix to extract the final shared key, thereby producing information-theoretically secure quantum keys for upper layers applications.

\section{One-time universal$_2$ hashing quantum digital signatures}\label{app:OTUHQDS}

\subsection{ Toeplitz hashing scheme}
The one-time universal hashing quantum digital signature (OTUH-QDS) protocol~\cite{2023-Yin,2023-Li} is built upon an almost XOR universal hash function, which we implement using a Linear-Feedback Shift Register-based Toeplitz hashing scheme. This hash function is defined by an $n \times m$ Toeplitz matrix, $H_{X,p}$, constructed from two $n$-bit secret parameters: a seed key $X$ distributed via the KGP in the quantum physical layer, and the coefficients of an irreducible polynomial $p$ generated locally. The first column of $H_{X,p}$ is the seed $X$, with subsequent columns generated recursively. The $n$-bit digest, $Dig$, of an $m$-bit message, $Mes$, is then computed as the matrix-vector product $Dig = H_{X,p} \cdot Mes \pmod 2$. 

\subsection{Detailed steps}

The protocol operates in two phases: distribution and messaging phases.

Distribution phase: The signer, forwarder, and verifier establish three sets of correlated $n$-bit keys via a KGP. These keys are denoted as $\{X_i\}$, $\{Y_i\}$, and $\{Z_i\}$, where the subscript $i \in \{s, f, v\}$ represents each party. The keys satisfy the correlation conditions: $X_{s} = X_{f} \oplus X_{v}$, $Y_{s} = Y_{f} \oplus Y_{v}$, and $Z_{s} = Z_{f} \oplus Z_{v}$. 

Messaging phase: The signing of a message $Mes$ is completed following the steps below.

\textit{1. Signing.} The signer generates a signature for a message $Mes$. First, the key $X_{s}$ is used as the seed, and a random irreducible polynomial, represented by its coefficients $p_{s}$, is generated locally. Together, these define the hash function $H_{X_{s},p_{s}}$. Subsequently, the message digest is calculated: $Dig = H_{X_{s},p_{s}}(Mes)$. Finally, the other two keys are used to encrypt the digest and the polynomial coefficients, yielding the signature $Sig = Dig \oplus Y_{s}$ and the encrypted polynomial $p = p_{s} \oplus Z_{s}$. 

\textit{2. Transmission and forwarding.} The signer transmits the triplet $\{Mes, Sig, p\}$ to the forwarder over an authenticated classical channel. Then, the forwarder transmits  $\{Mes, Sig, p\}$ and $\{  X_f, Y_f, Z_f \}$ to verifier for verification. 

\textit{3. Verification.} The verifier checks the message and signature first. If the verifier decides to accept, $\{  X_v, Y_v, Z_v \}$ will be sent to the forwarder for verification. The QDS succeeds if both the forwarder and the verifier decide to accept. 

The forwarder and the verifier perform an identical verification procedure. Each party independently reconstructs the signer's complete keys: $X_{s}, Y_{s}, Z_{s}$. Each party then executes the following steps: (i) Decrypt the received data using the reconstructed keys to recover the expected digest $Dig^{\prime} = Sig \oplus Y_{s}$ and the polynomial coefficients $p_{s}^{\prime} = p \oplus Z_{s}$. (ii) Construct the hash function using the reconstructed $X_{s}$ and $p_{s}^{\prime}$, and recompute the digest for the original message $Mes$: $Dig^{\prime \prime} = H_{X_{s},p_{s}\prime}(Mes)$. (iii) Compare the two digests. The signature is deemed valid if and only if $Dig\prime = Dig^{\prime \prime}$.

\subsection{Security analysis}

The security of the protocol relies on two core properties: unforgeability and non-repudiation. 

\textit{Unforgeability}: For an adversary without knowledge of the keys, the optimal strategy to forge a valid signature for a tampered message is equivalent to guessing the seed key $X_{s}$ of the hash function. The probability of a successful forgery, $\varepsilon_{\mathrm{for}}$, is bounded by $\varepsilon_{\mathrm{for}} \le m \cdot 2^{1-n}$, where $m$ and $n$ are the lengths of the message and the key, respectively. 

\textit{Non-repudiation}: Since the forwarder and the verifier can exchange their key shares over an authenticated channel, they can deterministically and independently reconstruct the complete set of keys and the exact hash function used by the signer. Consequently, they will always reach the same verification outcome, which prevents the signer from later denying the signature, thus achieving non-repudiation.

\section{Recursive QBA protocol}\label{app:QBA}
Our quantum consensus layer adopts the Recursive QBA~\cite{2023-Weng} as the consensus mechanism, making our blockchain naturally satisfy the two interactive consistency (\textbf{IC}) conditions: (1) All loyal replica nodes agree on the same order (\textbf{IC$_1$}); (2) If the main node is loyal, then all loyal replica nodes follow the loyal main node's order \textbf{IC$_2$}. Moreover, owing to the outstanding properties of the Recursive QBA, our blockchain also achieves almost one-half fault-tolerance, breaking the classical limit, and is fully decentralized.

\begin{figure*}[t!]
	\begin{center}
		\includegraphics[width=\textwidth]{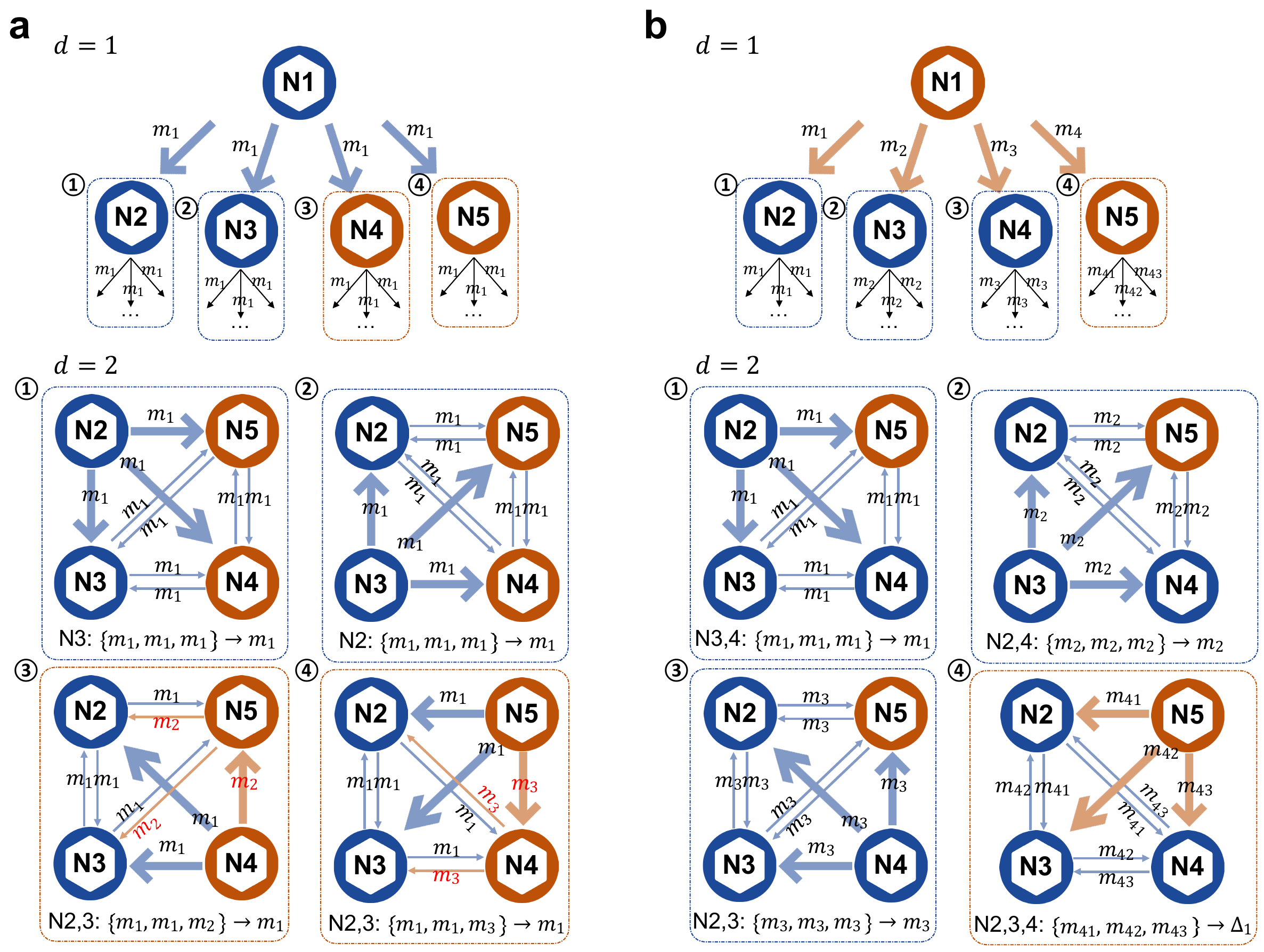}
	\end{center}
	\caption{\textbf{Schematic of recursive QBA with five nodes.} Node N1 is randomly selected as the main node, while N2-N5 act as replica nodes. (a) \textit{ N1 is loyal.} The final message lists obtained by honest nodes N2 and N3 are identical, $\{m_1, m_1, m_1, m_1\}$, enabling them to reach consensus on the correct command $m_1$ issued by N1.
		(b) \textit{ N1 is disloyal.} Even when the main node N1 sends inconsistent messages, the final message lists deduced by honest nodes N2, N3, and N4 remain consistent, $\{m_1, m_2, m_3, \Delta_1\}$, allowing them to still achieve consensus on the final output.}
	\label{fig:QBA-protocol}
\end{figure*}

The protocol reaches consensus among $N$ parties with up to $f$ malicious nodes. It has two main phases: a \textit{Broadcasting Phase} to share information and a \textit{Gathering Phase} to make decisions. Below, we first describe the general $N$-party protocol. Then, we explain the specific steps for the three-party ($N=3$) and five-party ($N=5$) cases. Fig.~\ref{fig:QBA-protocol} shows the five-party protocol when the main node is either honest or dishonest.

\subsection{N-party Recursive QBA}\label{DetailStepsGeneral}

The $N$-party recursive QBA protocol operates recursively across depths $d = 1, \dots, f$ to achieve consensus in the presence of up to $f$ malicious nodes.

\textit{Broadcasting Phase.} 
This phase disseminates messages through a recursive multicast structure. At depth $d=1$, the initial primary distributes a signed command to all replica nodes via a multicast round. For each subsequent depth $d$ (from 2 to $f$), every node that served as a backup at depth $d-1$ acts as a new primary. These nodes initiate new multicast rounds to redistribute the specific messages they received in the previous step to the remaining nodes. To ensure data integrity, an \textit{interactive consistency check} is performed at each step: a node is cryptographically restricted from forwarding a message at depth $d$ that differs from the one it acknowledged receiving at depth $d-1$. This mechanism forces malicious nodes to either forward consistent messages or be detected. Consequently, at the end of depth $f$, each honest node records a complete list of messages received through all valid forwarding paths.

\textit{Gathering Phase.} 
This phase resolves inconsistencies through a bottom-up majority-voting process. The procedure iterates from depth $d=f$ down to $1$. At the maximum depth $d=f$, each honest node initializes its gathering list using the locally recorded broadcasting list. For any depth $d < f$, the node constructs its gathering list by processing the data from depth $d+1$. Specifically, the value for a given entry in the list at depth $d$ is determined by taking the majority vote of the corresponding set of values received from the sub-primaries at depth $d+1$. This step effectively filters out incorrect values propagated by malicious forwarders at the lower level. Finally, at depth $d=1$, the node performs a majority vote on its top-level gathering list to produce the final consensus output.

\begin{figure*}[t!]
	\begin{center}
		\includegraphics[width=\textwidth]{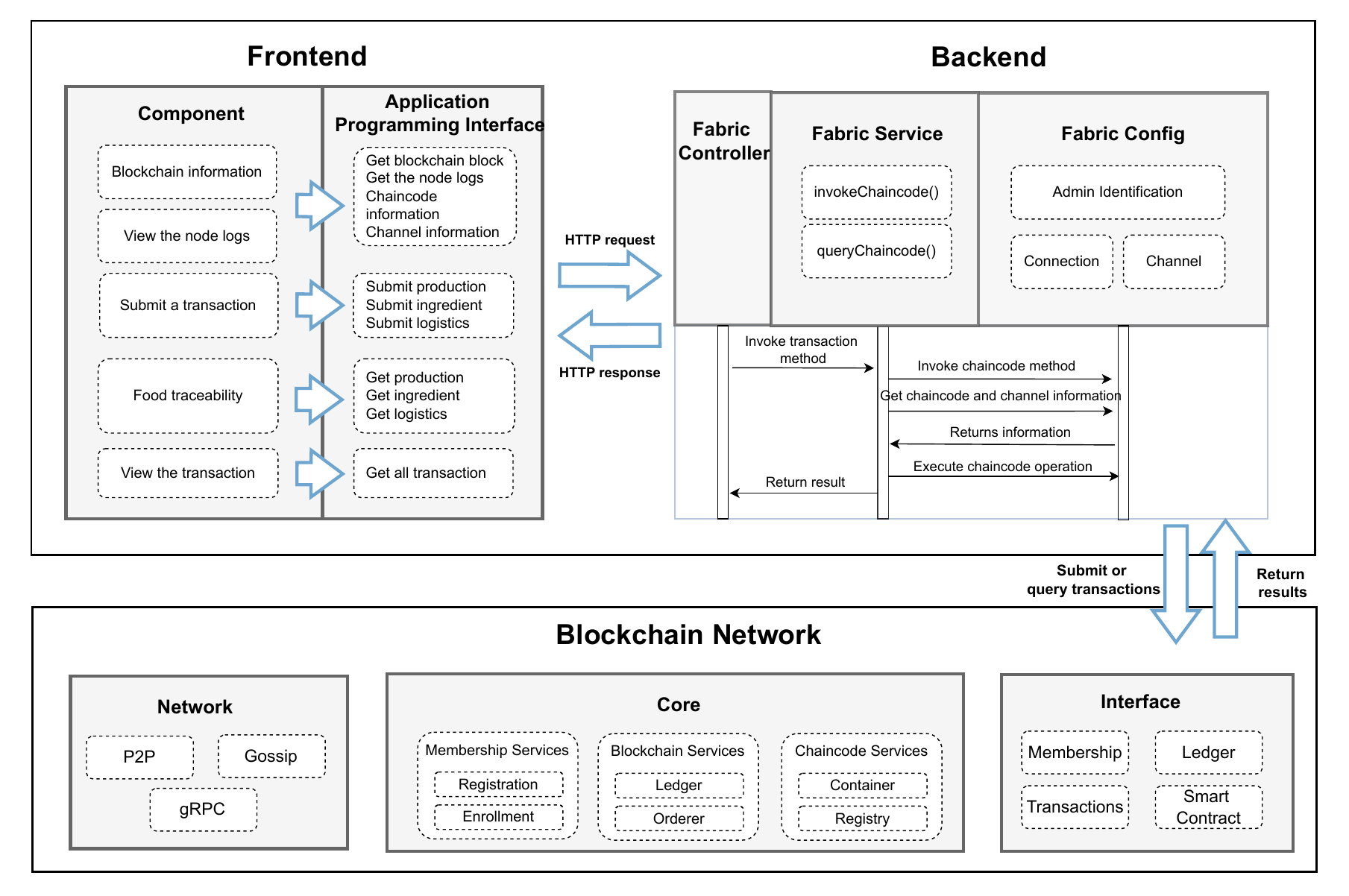}
	\end{center}
	\caption{\textbf{Interaction architecture of food traceability system.} The Vue-based frontend integrates user-facing components and associated Application Programming Interface (API) to collect and transmit product, ingredient, and logistics information to the Spring Boot microservice backend. The backend, comprising the Fabric Controller, Fabric Service, and Fabric Config modules, validates and normalizes incoming data, invokes or queries chaincode functions, and coordinates execution of the QBA consensus. The blockchain network, structured into Interface, Core, and Network layers, delivers membership management, immutable ledger updates, transaction ordering, smart contract execution, and peer-to-peer communication, ensuring transparent and tamper-proof traceability from data input to final result visualization.}
	\label{fig:Interaction_architecture}
\end{figure*}

\subsection{Performance and Security} 

\subsubsection{Fault tolerance}
The quantum consensus layer can tolerate up to $f$ malicious nodes out of $N$ participants, provided that 
\begin{equation}
    N \ge 2f+1. 
\end{equation}
This protocol breaks the classical one-third fault-tolerance limit because QDS provides asymmetric multiparty correlations. These correlations remove the classical assumption that communication channels between different pairs of nodes are independent.

\subsubsection{Communication complexity} 
The communication complexity $C$ is defined as the total number of three-party QDS operations required to complete a full consensus. It scales with the number of players $N$ and the number of malicious nodes $f$. The formula is given by:
\begin{equation}
\begin{aligned}
C  &= \mathrm{P}(N-1,2) + (N-1) \mathrm{P}(N-2,2)+\cdots\\
&\quad +(N-1)(N-2)\cdots(N-f+1) \mathrm{P}(N-f,2)\\
&= \sum_{m = 0}^{f-1}\mathrm{P}(N-1,2+m),
\end{aligned}
\label{complexity}
\end{equation}
where $\mathrm{P}(a,b) = a!/(a-b)!$ is the permutation function. The protocol requires a recursion depth equal to $f$. If the exact number of faulty nodes is unknown in practice, the recursion depth is set to the largest integer satisfying $N \ge 2f+1$, which is $\lfloor (N-1)/2 \rfloor$.

\subsubsection{Consensus rate}
To evaluate the performance of the consensus layer, we first determine the signature rate ($SR$) of its fundamental building block: the three-party OTUH-QDS. The $SR$ quantifies the number of signatures a specific triplet of nodes can generate per second. As each $n$-bit signature consumes $3n$ bits of secret key, the $SR$ is ultimately limited by the secure key rates ($R$) of the quantum links. In a QDS process with signer (node $i$), forwarder (node $j$), and verifier (node $k$), the performance is bottlenecked by the minimum key rate supplied to the verifier from the other two parties. Therefore, the signature rate for the triplet $(i,j,k)$ is given by
\begin{equation}
	SR_{i,j,k} = \min(R_{i,j}, R_{i,k})/3n.
\end{equation}
Here, the subscript in $R$ denotes the secure key rate of the quantum link between the two referred nodes. The rate is determined by the minimum of the two links involving the signer (node $i$) because the signer's keys are locally synthesized via XOR operation from the keys it shares with the other two parties, making the link between the verifier and forwarder non-critical for the rate calculation. With the signature rate for any given triplet established, we can now define the overall consensus rate ($CR$) for the entire network. The $CR$ is determined by the protocol's performance bottleneck, which is the minimum signature rate across all possible combinations of three-party participants, yielding the final consensus rate:
\begin{equation}
	CR = \mathop{\mathrm{min}}\limits_{ (i,j,k) \in T}(SR_{i,j,k})/C,
\end{equation}
where $T$ denotes all triplets consisting of three parties from all five nodes.

\subsubsection{Security analysis}
The security of the protocol rests on satisfying two Byzantine conditions: agreement among honest nodes (IC1) and validity of the honest primary's command (IC2). The total security failure probability, $\varepsilon_{\rm{tot}}$, is the sum of the failure probabilities of the underlying components: $\varepsilon_{\rm{tot}} = \varepsilon_{\rm{KGP}} + \varepsilon_{\rm{QBA}}$. Here, $\varepsilon_{\rm{KGP}}$ denotes the failure probability due to finite-size effects in the Quantum KGP~\cite{2017-Yin-MDI}. The term $\varepsilon_{\rm{QBA}}$ represents the probability of a forgery occurring in any of the $C$ signature instances. Using the union bound, this is estimated as
\begin{equation}
	\begin{aligned}
		\varepsilon_{\rm{QBA}} = 1-\left(  1-\varepsilon_{\rm{for}} \right )^C\approx C \cdot \varepsilon_{\rm{for}}
	\end{aligned}
	\label{varepsilonQBA}
\end{equation}
where the single-signature forgery probability is $\varepsilon_{\rm{for}} = m\cdot 2^{1-n}$ for a message of length $m$ and a key of length $n$~\cite{2023-Yin}.

\subsection{ Three-Party protocol demonstration ($N=3, f=1$)}\label{DetailSteps3}

We now describe the protocol flow for the three-party case, involving a main node (N1) and two replica nodes (N2, N3). With one potential malicious player ($f=1$), the protocol comprises a single recursive depth ($d=1$).

\textit{Case 1: The main node is Honest.} 

Here, N1 and N2 are honest, while N3 is malicious. In the \textit{Broadcasting Phase}, N1 initiates the multicast round ${MR}^1_{\text{N1}}$ with message $m^1_{\text{N1}} = m_1$. Due to the unforgeability property of QDS, the malicious N3 is prevented from successfully forging a different message when acting as the Forwarder. Consequently, the honest replica node N2 receives $m_1$ both directly from N1 and via forwarding from N3.

This results in N2's broadcasting list being $B^{1, \text{N2}}_{\text{N1}} = \{m_1, m_1\}$. In the subsequent \textit{Gathering Phase}, this list becomes the gathering list ($G^{1, \text{N2}}_{\text{N1}} = B^{1, \text{N2}}_{\text{N1}}$). The final output for N2 is therefore $m^{1, \text{N2}}_{\text{N1}} = \text{majority}(\{m_1, m_1\}) = m_1$, correctly matching the main node's order.

\textit{Case 2: The main node is Dishonest.}

In this scenario, N1 is malicious, while N2 and N3 are honest. The malicious main node attempts to equivocate during ${MR}^1_{\text{N1}}$ by sending conflicting messages: $m_1$ to N2 and $m_2$ to N3. When the honest replica nodes subsequently act as Forwarders for each other, they faithfully relay the message they received.

As a result, their respective broadcasting lists become $B^{1, \text{N2}}_{\text{N1}} = \{m_1, m_2\}$ and $B^{1, \text{N3}}_{\text{N1}} = \{m_2, m_1\}$. In the Gathering Phase, their gathering lists $G^{1, \text{N2}}_{\text{N1}}$ and $G^{1, \text{N3}}_{\text{N1}}$ are identical in content. Applying the deterministic majority function yields the same output for both:
$m^{1, \text{N2}}_{\text{N1}} = m^{1, \text{N3}}_{\text{N1}} = \text{majority}(\{m_1, m_2\}) = \Delta$.

\subsection{Five-Party demonstration ($N=5, f=2$)}\label{DetailSteps5}

We now outline the protocol execution for five players (N1, N2, N3, N4, N5) in the presence of up to two malicious actors ($f=2$). This scenario necessitates a two-depth recursion ($d=1, 2$). The illustration is provided in Fig.~\ref{fig:QBA-protocol}.

\begin{table*}[t!]
	\centering
	\caption{Experimental parameters and results. Loss denote the fiber transmission loss for the ten user pairs formed from five nodes (N1--N5). $\mu$ ($\nu$) represents the average photon number of the signal (decoy) states, and $P_{\mu}$ ($P_{\nu}$) represents the probability of selecting the average photon number $\mu$ ($\nu$). $P_{Z}$ ($P_{X}$) denotes the probability that the transmitter randomly chooses the $Z$ ($X$) basis. $n_{Z,\mu}$ ($n_{Z,\nu}$) denotes the total number of detection events when the transmitter prepares signal (decoy) states in the $Z$ basis and the receiver measures in the $Z$ basis, with $m_{Z,\mu}$ ($m_{Z,\nu}$) representing the corresponding number of detection errors. $n_{X,\mu}$ ($n_{X,\nu}$) and $m_{X,\mu}$ ($m_{X,\nu}$) denote the total detection events and corresponding errors when both parties select the $X$ basis and the transmitter prepares signal (decoy) states. $n_{Z}$ denotes the total $Z$-basis detection count after basis sifting, and $t$ denotes the time required to accumulate $n_{Z}$ events. $s_{Z,1}^{l}$ denotes the lower bound on single-photon detection events in the $Z$ basis; $E_{Z}$ denotes the $Z$-basis quantum bit error rate; $\phi_{Z}^{u}$ denotes the upper bound on the $Z$-basis phase error rate; $\lambda_{\rm EC}$ denotes the number of bits leaked during error correction; and $SKR$ denotes the achievable secure key rate for each link.}
	\begingroup
	\fontsize{9}{12}\selectfont 
	\setlength{\tabcolsep}{7pt} 
	\renewcommand{\arraystretch}{0.95} 
	\begin{tabular}{*{15}{c}}
			\\ \hline\hline
		\rule[-1ex]{0pt}{3.5ex}Link &  Loss(dB) &  $\mu$ &  $\nu$ &  $P_{\mu}$(\%) &  $P_{\nu}$(\%) &  $P_{Z}$(\%) &  $P_{X}$(\%) &  $n_{Z,\mu}$ &  $m_{Z,\mu}$ &  $n_{X,\mu}$ &  $m_{X,\mu}$  \\ \hline
		\rule[-1ex]{0pt}{3.5ex}N2-N1 &  0.7 &  0.54 &  0.15 &  77.91 &  22.09 &  94.01 &  5.99 &  9292211 &  69582 &  473854 &  6288  \\ \hline
		N3-N1 &  1.0 &  0.54 &  0.15 &  77.86 &  22.14 &  93.61 &  6.39 &  9329930 &  85889 &  484922 &  3698  \\ \hline
		N4-N1 &  20.0 &  0.53 &  0.14 &  77.70 &  22.30 &  94.35 &  5.65 &  9293901 &  133627 &  451003 &  3759  \\ \hline
		N5-N1 &  17.9 &  0.53 &  0.14 &  77.73 &  22.27 &  93.61 &  6.39 &  9291325 &  124747 &  415954 &  2888  \\ \hline
		N3-N2 &  1.4 &  0.54 &  0.15 &  77.75 &  22.25 &  94.00 &  6.00 &  9255128 &  95043 &  459203 &  3164  \\ \hline
		N4-N2 &  19.6 &  0.53 &  0.14 &  77.54 &  22.46 &  93.92 &  6.08 &  9287162 &  89051 &  439086 &  4080  \\ \hline
		N5-N2 &  18.0 &  0.53 &  0.14 &  77.73 &  22.27 &  93.94 &  6.06 &  9299845 &  95430 &  477487 &  4809  \\ \hline
		N4-N3 &  19.6 &  0.53 &  0.14 &  77.60 &  22.40 &  93.94 &  6.06 &  9281960 &  96446 &  447146 &  5223  \\ \hline
		N5-N3 &  16.4 &  0.53 &  0.14 &  77.63 &  22.37 &  93.92 &  6.08 &  9275980 &  102422 &  447513 &  5584  \\ \hline
		N5-N4 &  17.2 &  0.54 &  0.14 &  77.48 &  22.52 &  93.95 &  6.05 &  9297507 &  86291 &  329084 &  2619
		\\ \hline\hline
		\\ \hline\hline
		\rule[-1ex]{0pt}{3.5ex}Link &  $n_{Z,\nu}$ &  $m_{Z,\nu}$ &  $n_{X,\nu}$ &  $m_{X,\nu}$ &  $n_{Z}$ &  $t$(s) &  $s_{Z,1}^{l}$ &  $E_{Z}$(\%) &  $\phi_{Z}^{u}$(\%) &  $\lambda_{EC}$ &  SKR(kbps)  \\ \hline
		N2-N1 &  707789 &  4515 &  37210 &  607 &  $10^7$ &  12.1 &  5364553 &  0.74 &  3.94 &  760792 &  273.49  \\ \hline
		N3-N1 &  670070 &  9772 &  35396 &  359 &  $10^7$ &  12.3 &  4685924 &  0.96 &  2.73 &  944996 &  236.16  \\ \hline
		N4-N1 &  706099 &  12168 &  35114 &  334 &  $10^7$ &  992.5 &  5225023 &  1.46 &  2.60 &  1362079 &  2.98  \\ \hline
		N5-N1 &  708675 &  7780 &  32892 &  231 &  $10^7$ &  632.2 &  5340448 &  1.33 &  2.23 &  1251271 &  5.17  \\ \hline
		N3-N2 &  744872 &  12258 &  37165 &  291 &  $10^7$ &  14.1 &  5635941 &  1.07 &  2.06 &  1039132 &  267.40  \\ \hline
		N4-N2 &  712838 &  10103 &  34693 &  352 &  $10^7$ &  918.0 &  5263050 &  0.99 &  2.89 &  975435 &  3.59  \\ \hline
		N5-N2 &  700155 &  6398 &  36857 &  480 &  $10^7$ &  636.7 &  5279193 &  1.02 &  3.06 &  996110 &  5.09  \\ \hline
		N4-N3 &  718040 &  7593 &  35934 &  571 &  $10^7$ &  921.9 &  5422257 &  1.04 &  3.28 &  1013671 &  3.56  \\ \hline
		N5-N3 &  724020 &  6782 &  35689 &  408 &  $10^7$ &  443.0 &  5531786 &  1.09 &  3.78 &  1056054 &  7.20  \\ \hline
		N5-N4 &  702493 &  5918 &  25463 &  254 &  $10^7$ &  519.4 &  5069154 &  0.92 &  2.89 &  917569 &  6.15  \\ \hline\hline
	\end{tabular}
	\endgroup
	\label{table_rawdate}
\end{table*}

\begin{table*}[t]
	\centering
	\caption{\textbf{Blockchain performance evaluation results.} Performance metrics of different blockchain transactions measured using the benchmarking tool Hyperledger Caliper. The evaluated functions correspond to distinct supply chain lifecycle stages: CreateFoodItems (initializing product specifications), AddIngredientInfo (recording ingredient provenance), AddLogisticsInfo (updating logistics records), and QueryCompleteInfo (retrieving the full chain of custody). Succeeded and Failed indicate the number of successfully and unsuccessfully processed transactions, respectively. Send Rate (transactions per second) refers to the rate at which transactions are sent to the blockchain network, Throughput (transactions per second) denotes the rate of transactions successfully processed and committed, and Average Latency (s) represents the average transaction confirmation time, with minimum and maximum values shown in parentheses. These metrics collectively reflect the system's processing efficiency, responsiveness, and resource utilization under different load conditions.}
	\label{tab:perf_eval_updated}
	
	\renewcommand{\arraystretch}{1.5}
	\setlength{\tabcolsep}{4pt} 
	
	\begin{tabular*}{\textwidth}{@{\extracolsep{\fill}}lccccc@{}}
		\toprule
		\textbf{Transaction Name} & \textbf{Succeed} & \textbf{Failed} & 
		\textbf{\makecell{Send Rate\\(transactions per second)}} & 
		\textbf{\makecell{Average Latency (s)\\(min--max)}} & 
		\textbf{\makecell{Throughput\\(transactions per second)}} \\
		\midrule
		
		\multirow{6}{*}{CreateFoodItems}
		& 1000  & 0 & 100.1  & 0.88 (0.51--1.22) & 90.9 \\
		& 3000  & 0 & 300.2  & 1.41 (0.55--2.57) & 253.1 \\
		& 5000  & 0 & 500.4  & 2.43 (0.56--3.54) & 396.2 \\
		& 7000  & 0 & 700.5  & 4.91 (0.63--7.87) & 463.9 \\
		& 9000  & 0 & 900.4  & 7.92 (0.77--11.62) & 476.4 \\
		& 10000 & 0 & 998.1  & 8.66 (0.77--11.71) & 513.7 \\[5pt]
		
		\multirow{6}{*}{AddIngredientInfo}
		& 1000  & 0 & 100.1  & 0.90 (0.51--2.07) & 83.6 \\
		& 3000  & 0 & 300.2  & 1.32 (0.52--2.15) & 251.3 \\
		& 5000  & 0 & 500.4  & 2.34 (0.54--3.50) & 405.3 \\
		& 7000  & 0 & 700.2  & 4.51 (0.58--7.23) & 462.0 \\
		& 9000  & 0 & 900.3  & 8.23 (0.60--12.32) & 467.7 \\
		& 10000 & 0 & 999.8  & 8.68 (0.66--12.79) & 502.8 \\[5pt]
		
		\multirow{6}{*}{AddLogisticsInfo}
		& 1000  & 0 & 100.1  & 0.90 (0.51--2.08) & 83.3 \\
		& 3000  & 0 & 300.2  & 1.34 (0.52--1.84) & 274.2 \\
		& 5000  & 0 & 499.6  & 2.34 (0.52--3.48) & 416.8 \\
		& 7000  & 0 & 700.2  & 4.33 (0.57--6.95) & 478.4 \\
		& 9000  & 0 & 899.6  & 6.96 (0.62--10.85) & 518.0 \\
		& 10000 & 0 & 1000.2 & 10.20 (0.67--14.76) & 444.8 \\[5pt]
		
		\multirow{6}{*}{QueryCompleteInfo}
		& 1000  & 0 & 100.1  & 0.89 (0.50--1.29) & 91.0 \\
		& 3000  & 0 & 300.2  & 1.43 (0.52--2.27) & 253.5 \\
		& 5000  & 0 & 500.4  & 2.44 (0.54--3.65) & 399.2 \\
		& 7000  & 0 & 700.4  & 4.88 (0.64--7.51) & 473.6 \\
		& 9000  & 0 & 900.3  & 7.60 (0.62--12.09) & 506.1 \\
		& 10000 & 0 & 999.9  & 12.61 (0.69--16.73) & 403.8 \\
		\bottomrule
	\end{tabular*}
	
	\vspace{2pt} 
\end{table*}

\textit{Case 1: The main node is Honest.}

Consider the scenario where the main node N1 and replica nodes N2, N3 are honest, while N4 and N5 are malicious. In the \textit{Broadcasting Phase}, the honest main node N1 initiates ${MR}^1_{\text{N1}}$ with the message $m^1_{\text{N1}} = m_1$. Owing to the unforgeability of QDS, all honest players consistently record $m_1$, resulting in identical broadcasting lists, e.g., $B^{1, \text{Ni}}_{\text{N1}} = \{m_1, m_1, m_1, m_1\}$ for $i\in \{2,3,4,5\}$. While N4 and N5 may attempt to collude during the QDS process with N1, such actions are inconsequential, as the internal lists of malicious players do not affect the consensus among honest nodes. At depth $d=2$, the secondary rounds ${MR}^2_{\text{N1} \to \text{Ni}}$ commence. When the current primary is honest ($i\in\{2,3\}$), they multicast $m_1$, yielding the broadcasting lists $B^{2, \text{Ni}}_{\text{N1} \to \text{N2}} = B^{2, \text{Ni}}_{\text{N1} \to \text{N3}}= \{m_1, m_1, m_1\}$. Conversely, when the primary is malicious ($i\in\{4,5\}$), they can inject conflicting messages (e.g., $m_2$ and $m_3$) only through collusion. This leads to the divergent lists $B^{2, \text{Ni}}_{\text{N1} \to \text{N4}} = \{m_1, m_1, m_2\}$ and $B^{2, \text{Ni}}_{\text{N1} \to \text{N5}} = \{m_1, m_1, m_3\}$.

In the \textit{Gathering Phase}, each honest player initiates the process at $d=2$ by equating their gathering lists to their broadcasting lists. Subsequently, they compute a deduced message for each sub-branch via a majority vote. For instance, $m^{2, \text{N2}}_{\text{N1} \to \text{N4}} = \text{majority}(\{m_1, m_1, m_2\}) = m_1$. This step effectively filters out malicious attempts, revealing that the message originally received by every honest replica from N1 was indeed $m_1$. These deduced messages form the top-level gathering list, $G^{1, \text{N2}}_{\text{N1}} = \{m_1, m_1, m_1, m_1\}$. The final output is $m^{1, \text{N2}}_{\text{N1}} = \text{majority}(G^{1, \text{N2}}_{\text{N1}}) = m_1$, thereby satisfying conditions \textbf{IC$_1$} and \textbf{IC$_2$}.

\textit{Case 2: The main node is Dishonest.}

Next, consider the case where N1 and N5 are malicious, while N2, N3, and N4 remain honest. In the \textit{Broadcasting Phase}, during ${MR}^1_{\text{N1}}$, N1 sends conflicting messages ($m_1, m_2, m_3$) to the honest replica nodes. Furthermore, when N5 acts as the forwarder, the forwarded messages may differ across QDS instances. This results in discrepant broadcasting lists, e.g., $B^{1, \text{Ni}}_{\text{N1}} = \{m_1, m_2, m_3, m_{4(i-1)}\}$ for $i\in \{2,3,4\}$. At depth $d=2$, the unforgeability of QDS ensures that N5 cannot forge new messages, and the inherent consistency check constrains N5 to multicasting the specific message versions previously acknowledged by the recipients. As a result, every honest player records the exact same set of messages during ${MR}^2_{\text{N1} \to \text{N5}}$. Their broadcasting lists will therefore be identical:
\[
B^{2, \text{N2}}_{\text{N1} \to \text{N5}} = B^{2, \text{N3}}_{\text{N1} \to \text{N5}} = B^{2, \text{N4}}_{\text{N1} \to \text{N5}} = \{m_{41}, m_{42}, m_{43}\}
\]

In the \textit{Gathering Phase}, all honest players derive an identical deduced value from these broadcasting lists by applying the deterministic majority function to the same input set. We define this common result as $\Delta_1 = \text{majority}(\{m_{41}, m_{42}, m_{43}\})$. This determinism ensures that when honest players construct their top-level gathering lists, the content is uniform. Each node independently concludes that the messages effectively received from N1 by N2, N3, N4, and N5 were $m_1, m_2, m_3$, and $\Delta_1$, respectively. This yields identical top-level gathering lists for all honest players:
\[
G^{1, \text{N2}}_{\text{N1}} = G^{1, \text{N3}}_{\text{N1}} = G^{1, \text{N4}}_{\text{N1}} = \{m_1, m_2, m_3, \Delta_1\}
\]
Ultimately, a final majority vote on this shared set of values produces the identical output $\Delta_2$, thereby satisfying condition \textbf{IC$_1$}.

Applying the above theoretical framework to our five-party experiment, we set the total number of nodes to $N=5$ to tolerate up to $f=2$ malicious nodes. This results in a communication complexity of $C=36$. We set the signature length to $n=64$ bits. Across our ten point-to-point quantum links, the minimum secure key rate is 2.98 kbps. Based on these link rates, the signature rate ($SR$) for any three-node group is at least 15.51 times per second (tps), with an average of 744.10 tps and a maximum of 1424.42 tps. By dividing the minimum $SR$ by the complexity $C=36$, our system achieves a final consensus rate of $CR \approx 0.43$ tps. Under these parameters, the quantum consensus layer guarantees a security level of approximately $10^{-11.4}$.

\section{Food traceability system}\label{app:FTS}
Fig.~\ref{fig:Interaction_architecture} illustrates the end-to-end architecture of the proposed food traceability system, detailing the coordinated interactions between the user-centric frontend, the microservice-based backend, and the quantum blockchain network. This architectural layout maps directly to the six-stage operational flow presented in Fig. 1 of the main text, ensuring a coherent alignment between the conceptual framework and its technical implementation.

The operational cycle initiates at the Vue-based frontend, where users input food-related data, including product specifications, ingredient provenance, and logistics records. User actions trigger the component modules responsible for transaction submission, data viewing, and traceability queries, while the associated API layer dispatches structured HTTP requests to the backend service. This stage reflects the "Input food data" and "Send to backend service" phases in Fig. 3 of the main text, ensuring that raw supply chain information is seamlessly transferred to the processing core.

Upon receiving the requests, the Spring Boot backend, configured through the Fabric Controller, Fabric Service, and Fabric Config modules, validates the data, performs normalization and preliminary verification, and maps each operation to the corresponding chaincode function. Invocations for transaction execution (via invokeChaincode) and data retrieval (via queryChaincode) are directed to the blockchain interface, after which the orderer nodes initiate the quantum Byzantine agreement (QBA) consensus procedure. This constitutes the "Execution quantum consensus" stage, where transaction ordering is resolved across participating nodes, guaranteeing immutability and fault tolerance.

Once consensus is achieved, the process culminates in the "Record on blockchain" phase. Here, the Core layer commits verified transactions to the ledger, leveraging Membership Services and chaincode management to ensure the data remains authoritative and tamper-resistant. Finally, to close the loop, authorized stakeholders perform the "Query traceability result" and "View result" phases. By accessing backend APIs, users retrieve the complete immutable history-including block, transaction, and node logs-which is rendered by the frontend to visualize the temporal chain of custody, thereby enhancing supply chain transparency and consumer confidence.

\section{Detailed experimental results of the quantum physical layer}\label{app:TS1}
Table~\ref{table_rawdate} shows the detailed experimental results of the metropolitan quantum network based on silicon photonic integrated circuits.

\section{Benchmarking results for the food traceability system}\label{app:TS2}
Table~\ref{tab:perf_eval_updated} presents detailed performance metrics for a quantum-blockchain-based food traceability system.


\bibliography{Chip-DC-QBC}
\bibliographystyle{naturemag}

\end{document}